\documentclass[preprint,12pt]{elsarticle}

\usepackage{graphics}
\usepackage{graphicx}
\usepackage{epsfig}

\usepackage{amssymb}

\journal{Solid State Communications}

\begin{document}

\begin{frontmatter}

\title{Role of spin orbit coupling and electron correlation in the
electronic structure of a 5$d$ pyrochlore, Y$_2$Ir$_2$O$_7$}

\author{Kalobaran Maiti}

\ead{kbmaiti@tifr.res.in}

\address{Department of Condensed Matter Physics and Materials
Science, Tata Institute of Fundamental Research, Homi Bhabha Road,
Colaba, Mumbai - 400 005, INDIA}

\begin{abstract}
Spin orbit coupling in 5$d$ transition metal oxides such as Ir
oxides is expected to be strong due to large atomic number of Ir and
electron correlation strength will be weak due to large radial
extension of the 5$d$ orbitals. Hence, various anomalous electronic
properties often observed in these systems are attributed to large
spin-orbit interaction strength. Employing first principles
approaches, we studied the electronic structure of Y$_2$Ir$_2$O$_7$,
which is insulating and exhibits ferromagnetic phase below 150 K.
The calculated results reveal breakdown of both the above paradigms.
The role of spin-orbit interaction is found to be marginal in
determining the insulating ground state of Y$_2$Ir$_2$O$_7$. A large
electron correlation strength is required to derive the experimental
bulk spectrum.
\end{abstract}


\end{frontmatter}

\section{Introduction}

Spin orbit interaction in an atom is the interaction of electronic
spin moment with the magnetic field experienced by the moving
electron in the electric field of the nucleus. Since the nuclear
charge enhances with the increase in atomic number of the elements,
the heavier atoms will have larger spin-orbit coupling strength,
$\zeta$. On the other hand, the electron-electron Coulomb repulsion
strength, $U$ is inversely related to the spatial separation between
the electrons. Therefore, $U$ will reduce with the enhancement of
the radial extension of the electronic states. Based on these
paradigms, various anomalous properties observed in compounds
containing heavier elements are often attributed to large $\zeta$.

The scenario in solids can be significantly different due to the
finite overlap of the electronic states centered at different
lattice sites. The electrons can move among various lattice sites;
the electronic states are extended beyond the atomic limit. While
the electron correlation leads to localization of the electronic
states (essentially an on-site parameter), the hopping interaction
strength, $t$ represents the itineracy of the electrons which is
manifested in the bandwidth. Therefore, the parameters those
determine the electronic ground state are $\zeta$, $U$ and $t$. It
is clear that in addition to the correlation induced localization
effect, the degree of local character of the electronic states in
solids depends strongly on the underlying crystallographic structure
(band structure effect). It is thus important to investigate the
influence of such crystal structure induced effects on $\zeta$ and
$U$ in a solid. In this paper, we show evidences of the deviation
from both the paradigms for heavier elements mentioned above in a
pyrochlore, Y$_2$Ir$_2$O$_7$.

The compounds in pyrochlore structure, in general, have drawn
significant attention in the recent times followed by the discovery
of various exotic phenomena such as spin ice behavior,\cite{SpinIce}
superconductivity,\cite{Cd2Re2O7-1} correlation induced metal
insulator transitions\cite{R2Mo2O7} {\em etc}. Among Ir based
compounds, Pr$_2$Ir$_2$O$_7$ is a metal and shows spin-liquid
behavior,\cite{SLpr2ir2o7} anomalous Hall effect\cite{AHEpr2ir2o7}
{\em etc}. Interestingly, Y$_2$Ir$_2$O$_7$ is predicted to be a {\em
Mott insulator}, which exhibits a weak ferromagnetic transition at
around 150~K.\cite{Fukazawa,soda,fukazawa1} A high resolution
photoemission study reveals signature of finite density of states at
the Fermi level although it is an insulator\cite{y2ir2o7} as also
observed in an another Ir based compound, BaIrO$_3$.\cite{bairo3}
$U$ is expected to be small in these compounds due to the large
radial extension of the Ir 5$d$ orbitals, which questions the
feasibility of Mott insulating phase in these
systems.\cite{allen,Cd2Re2O7-2,bairo3band} A recent study on another
Ir-based compound, Sr$_2$IrO$_4$ suggests importance of spin orbit
coupling in deriving the insulating ground state despite weak
electron correlation strength.\cite{sr2iro4} Thus, Y$_2$Ir$_2$O$_7$
is an unique test case to study the role of spin orbit coupling and
electron correlation in deriving unusual ground state properties.

\section{Calculational details}

In order to address the issue, we calculated the electronic band
structure of Y$_2$Ir$_2$O$_7$ using state-of-the-art full potential
linearized augmented plane wave method (WIEN2k software)\cite{wien}
within the local spin density approximations, LSDA. The convergence
for different calculations were achieved considering 512 $k$ points
within the first Brillouin zone. The error bar for the energy
convergence was set to $<$~0.1~meV per formula unit (fu). In every
case, the charge convergence was achieved to be less than 10$^{-3}$
electronic charge. Structural parameters used for these calculations
were estimated via Rietveld refinement of the $x$-ray diffraction
pattern.\cite{fukazawa1,ravi-thesis}

Y$_2$Ir$_2$O$_7$ forms in cubic pyrochlore structure with the
lattice constant, $a$~=~10.20~\AA; space group $Fd{\bar{3}}m$. The
atomic positions are; Y: 16d, Ir: 16c, O1: 48f ($x$ = 0.35461) and
O2: 8b. The muffin-tin radii ($R_{MT}$) for Y, Ir, O1 and O2 were
set to 2.2~a.u., 2.1~a.u., 1.85~a.u. and 1.85~a.u., respectively.

\section{Results and Discussions}

LSDA calculations are widely known to capture the magnetic phase in
various 3$d$, 4$d$ and 5$d$ transition metal oxides
well.\cite{bairo3band,davidsingh,ruth,ddprl} However, the energetics
in the present case is somewhat different. The ground state energy
corresponding to the non-magnetic solutions is about 3.7 meV/fu
lower than that for the spin-polarized calculations. This suggests
that simple Stoner like criteria may not be enough to capture the
experimentally observed ferromagnetic ground state. This is
consistent with the conclusion in a recent photoemission study
revealing $|\epsilon - \epsilon_F|^{1.5}$ dependence of the high
resolution photoemission lineshape: electron-magnon coupling may be
important.\cite{y2ir2o7}

The magnetic moment centered at Ir and O1 sites are about
0.27~$\mu_B$ and 0.04~$\mu_B$ respectively. The total moment of
0.9~$\mu_B$/fu is close to the saturation moment of 1 $\mu_B$
expected for $S$ = 1/2 state of Ir$^{4+}$. However, this is
significantly high considering that no saturation was observed
experimentally even at a high field of 12
Tesla.\cite{fukazawa1,ravi-thesis} The magnetic moment in the
paramagnetic phase was found to be about 0.55~$\mu_B$ ($\mu =
2\sqrt{S(S+1)} \mu_B$ assuming spin only value), which corresponds
to $S$ = 0.07. This is about 14\% of the $S$=1/2. Such a small
paramagnetic moment indicate that the small magnetic moment is
intrinsic to this system and not due to antiferromagnetic exchange
interactions. Moreover, the Curie-Weiss temperature, $\theta_p$
estimated from the paramagnetic phase is positive. All these results
and other studies based on specific heat and magnetization
measurements rule out the possibility of canted antiferromagnetism
as the reason for small magnetic moment. The other possibility for
the small magnetic moment in these Ir-compounds can be due to spin
glass ordering that cannot be calculated using such methods. In any
case, it is clear that these LSDA results overestimates the magnetic
moment.

It is not possible to capture a spin-glass phase using these ab
initio band structure calculations. Moreover, the crystal structure
is geometrically frustrated. Considering the facts that all the bulk
studies suggest ferromagnetic ordering in this compound and the
energy distribution of the density of states in the ferromagnetic
phase is close to that found in the non-magnetic phase, we have
studied the evolution of the electronic structure corresponding to
the ferromagnetic phase with spin-orbit coupling and electron
correlation in the following part. We also compare these results
with the experimental electronic structure.

\begin{figure}
\vspace{-2ex}
\begin{center}
\includegraphics[width=0.5\textwidth]{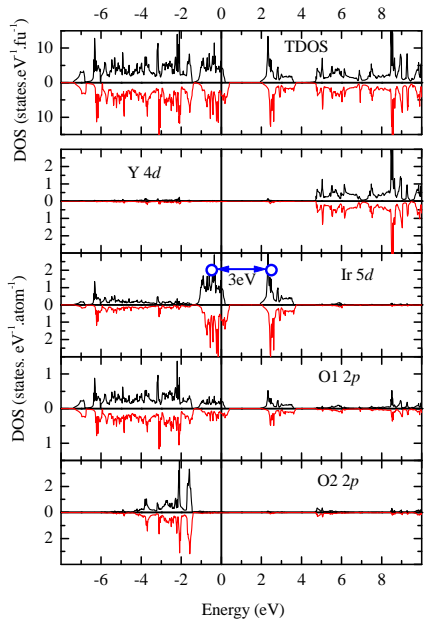}
\vspace{-8ex}
\end{center}
\caption{(Color online) Spin polarized total density of states
(TDOS), Y 4$d$ partial density of states (PDOS), Ir 5$d$ PDOS, O1
2$p$ PDOS and O2 2$p$ PDOS obtained from LSDA calculations. The up
and down spin density of states are represented by positive and
negative $y$-axis.}
\end{figure}

In Fig. 1, we show the density of states (DOS) calculated for the
ferromagnetic configuration of all the moments. Y 4$d$ partial DOS
(PDOS) appear primarily above 4~eV with negligible contribution
below the Fermi level, $\epsilon_F$. There are two groups of
features in the occupied part of the total DOS (TDOS). The energy
range, -7.5 to -1.5~eV is dominated by O1 2$p$ and O2 2$p$ PDOS with
a small contribution from Ir 5$d$ PDOS. O2 2$p$ PDOS appear between
-5.0 to -1.5 eV, where a small contribution from Y 4$d$ states are
also seen. This suggests existence of finite Y-O2 covalency. O1 2$p$
- Ir 5$d$ bonding states with dominant O1 2$p$ character appear
below -4 eV. The antibonding bands having essentially Ir 5$d$
character appear above -1.2~eV energies. Since the oxygens
surrounding the Ir sites form octahedral symmetry, the Ir 5$d$
levels split into a triply degenerate $t_{2g}$ band appearing
between -1.2~eV and 0.3~eV energies. The doubly degenerate $e_g$
bands appear above 2~eV energies. Clearly, the crystal field
splitting is significantly strong as expected for 5$d$ orbitals;
large radial extension of 5$d$ orbitals leads to a greater exposure
to the crystal field. One can estimate the crystal field splitting
by calculating the separation of the center of mass of the $t_{2g}$
and $e_g$ bands. The calculated center of mass of the $t_{2g}$ and
$e_g$ bands shown by open circles in the figure. The energy
separation of these two points is found to be 3~eV which is quite
large compared to the values found in 3$d$ transition metal oxides.

TDOS is large at $\epsilon_F$, suggesting a metallic phase. The
eigen energies for the up and down spin DOS is found to be similar
in every case except Ir 5$d$ states. The exchange splitting
estimated by the relative shift of the up and down spin Ir 5$d$
density of states is about 0.23 eV.

In order to investigate the role of spin-orbit coupling in the
electronic structure, we compare the results obtained in the
presence and absence of spin-orbit coupling. The magnetic field was
applied along (001) direction for all the calculations. In the
presence of spin-orbit coupling, the total energy corresponding to
the ferromagnetic solution is still higher by about 34 meV/fu than
that for the non-magnetic solution. The magnetic moment centered at
various lattice sites becomes much smaller than the values obtained
without spin-orbit coupling. The moments centered at Ir and O1 sites
are 0.044$\mu_B$ and 0.006 $\mu_B$. The total moment is found to be
about 0.137 $\mu_B$/fu. All these numbers are close to the
experimentally observed value corresponding to $S$ =
0.07.\cite{fukazawa1,ravi-thesis}

\begin{figure}
\vspace{-2ex}
\begin{center}
\includegraphics[angle=0,width=0.5\textwidth]{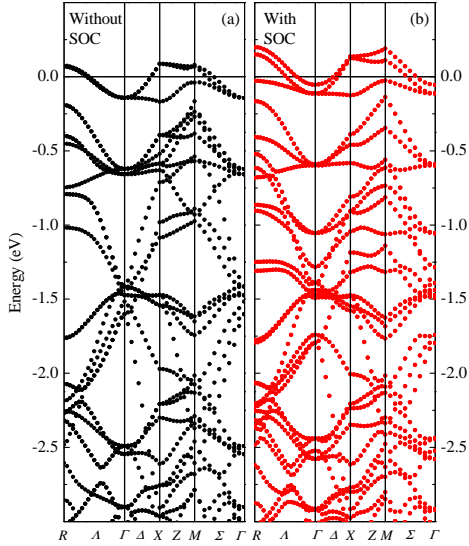}
\vspace{-12ex}
\end{center}
\caption{(Color online) Energy band dispersion obtained for the
non-magnetic calculations (a) without spin orbit coupling and (b)
with spin orbit coupling.}
\end{figure}

The calculated band dispersions are shown in Fig. 2. In order to
have better clarity in the figure, we have compared the band
dispersions for the non-magnetic solutions. The $t_{2g}$ bands
appearing in the vicinity of $\epsilon_F$ are significantly narrow
although the Ir 5$d$ orbitals have large radial extension. The
dispersions of the energy bands at lower energies become larger.
Evidently, the degeneracy of the bands are lifted due to the
inclusion of spin-orbit coupling as expected. One band of the triply
degenerate $t_{2g}$ band shifts completely below the Fermi level and
exhibit small narrowing. The other two bands crossing $\epsilon_F$
becomes wider when spin-orbit coupling is considered.

\begin{figure}
\vspace{-2ex}
\begin{center}
\includegraphics[angle=0,width=0.5\textwidth]{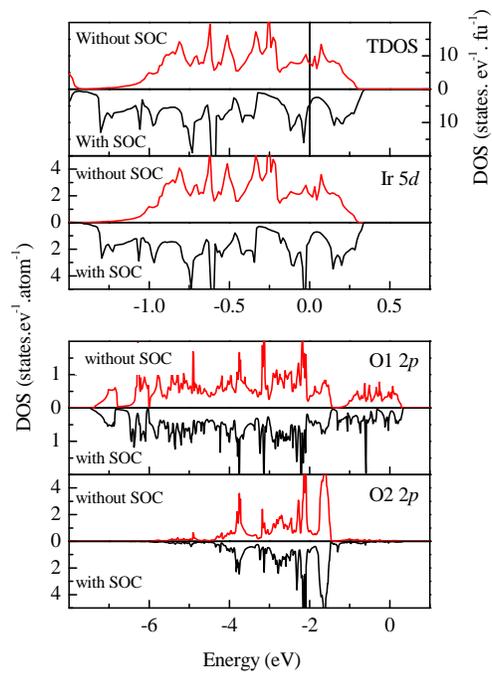}
\vspace{-4ex}
\end{center}
\caption{(Color online) Comparison of the TDOS, Ir 5$d$, O1 2$p$ and
O2 2$p$ density of states obtained from the non-magnetic
calculations with and without spin-orbit interactions.}
\end{figure}

In Fig. 3, we show the TDOS contributions and various partial DOS
corresponding to Ir 5$d$, O1 2$p$ and O2 2$p$ orbitals. Clearly the
character of the density of states appearing in the whole energy
range shown remain quite close in both the calculations.
Specifically, O1 2$p$ and O2 2$p$ partial DOS remain almost
insensitive to the consideration of the spin-orbit coupling as
expected. The inclusion of spin-orbit coupling leads to an
enhancement of overall width of the $t_{2g}$ band as also seen in
Fig. 2. The relative shift of the energy bands observed in Fig. 2 is
manifested by a small dip at $\epsilon_F$.

The above results can be explained as follows: the band width is
essentially determined by the hopping interaction strength, $t$,
$$t_{ij} = <\psi_j\mid H\mid\psi_i>$$ where $i$ and $j$ are the
site indices. The value of $t$ depends on the hybridization of the
electronic states centered at $i^{th}$ and $j^{th}$ sites. On the
other hand, spin-orbit coupling strength can be written as
$\epsilon(SOC) = \zeta S.L$ ($S$ and $L$ are the spin and orbital
moment of the electron). The spin-orbit coupling strength, $\zeta$
can be expressed as,
$$\zeta = {{2\mu_B}\over{\hbar m_e e c^2}}.{1\over r}
{{\partial V}\over{\partial r}}$$ where, $V$ is the potential energy
of electron in the central field of the nucleus. The magnitude of
$\zeta$ depends on the atomic number and is essentially an on-site
parameter. If $t$ is zero, the behavior will be close to the atomic
case. For finite $t$, the influence of the core potentials on the
conduction electrons will be less than the atomic case as such core
potentials will be seen by conduction electrons for lesser time.
Evidently, $\zeta$ will not have significant influence on $t$ unless
$\zeta$ is very large leading to a large change in energy difference
between the Ir 5$d$ and O 2$p$ eigen energies, which in turn will
influence the overlap integral of the 5$d$ orbitals with the ligand
orbitals.

In the present case, a close look suggests that among three bands in
the triply degenerate $t_{2g}$ band, one band is slightly narrowed
upon consideration of spin-orbit coupling. This energy band (see
Fig. 2) is completely occupied and appear below $\epsilon_F$. Since
the electronic properties are essentially determined by the energy
bands crossing the Fermi level, this band narrowing will have
negligible influence in determining the electronic properties of
this compound. The other two bands crosses $\epsilon_F$ and exhibit
a small enhancement in width. Thus, the spin-orbit coupling induced
change in bandwidth does not suggest any qualitative change in the
electronic properties of this system.

Apart from the above discussion involving influence in the
bandwidth, the spin-orbit coupling induced degeneracy lifting can
still play an important role in determining the influence of
electron correlation in the electronic properties. This has been
addressed by several authors in the past. Solving a $D$-fold
degenerate Hubbard model using the Gutzwiller approximation as well
as slave-boson approach, it was observed that the critical value of
$U$ (= $U_c$) required for metal-insulator transition increases
monotonically with the band filling, $n$. $U_c$ is maximum at half
filling,\cite{Lu} where $U_c \sim (D + 1)W$. The calculations based
on dynamical mean field theory\cite{kotliar} suggested $U_c = 1.85
\times W$. Exact calculations\cite{gunnarsson} at half filling also
suggest $U_c = \sqrt{D} W$. Consideration of inter-band
hopping\cite{Priya-DD} indicated significantly more pronounced
dependence of $U_c$ on the degeneracy. It is apparent that the
lifting of degeneracy due to spin orbit coupling will push $U_c$
towards smaller values. Thus, the spin-orbit coupling may in turn
lead to metal-insulator transition in weakly correlated system.

Being examined the influence of spin-orbit coupling in this system,
we now turn towards the electron correlation induced effects. In all
the following calculations, spin-orbit coupling is included. The Ir
5$d$ moments show interesting evolution with the increase in $U$.
The orbital and spin magnetic moments corresponding to Ir 5$d$
states are found to be 0.19 $\mu_B$ \& 0.25 $\mu_B$ for $U$ = 2 eV,
0.48 $\mu_B$ \& 0.5 $\mu_B$ for $U$ = 4 eV, and 0.53 $\mu_B$ \& 0.53
$\mu_B$ for $U$ = 6 eV. Evidently, both the spin and orbital
magnetic moment enhances with the increase in $U$ as observed in
other systems.\cite{andersen} The enhancement of the orbital
magnetic moment is more pronounced than the spin moment. Such
enhancement can be attributed to the enhancement of the local
character of the electronic states due to electron correlation. The
orbital moment is significantly large in this systems compared to
the values found in 3$d$ and 4$d$ transition metal oxides. Among all
the numbers found for different values of $U$, the magnetic moment
corresponding to $U$ = 2 eV is found to be the lowest and closer to
the experimental results. The calculated magnetic moment is always
found to be larger than the experimental value.

The calculated DOS for different values of $U$ including spin orbit
coupling are shown in Fig. 4. The DOS corresponding to the up and
down spin channels are shown along regular and reversed y-axis,
respectively. No hard gap is observed for $U$ as large as 6 eV. The
calculated results reveal many interesting features. Firstly, the
near negligible Ir 5$d$ contributions observed below -2 eV energy
range in LSDA results (see Fig. 1) becomes more and more significant
with the increase in $U$. This means, the Ir 5$d$ character enhances
in the bonding bands and subsequently, relative O 2$p$ character in
the antibonding band appearing near $\epsilon_F$ enhances. Thus, the
Ir 5$d$-O 2$p$ covalency is significantly influenced by the electron
correlation.

Secondly, the up spin feature in the vicinity of $\epsilon_F$
(antibonding band having primarily Ir 5$d$ character) gradually
moves towards lower energies and have marginal contributions at
$\epsilon_F$ for $U \geq$~4 eV. Distinct signatures of the coherent
feature (delocalized electronic states in the vicinity of
$\epsilon_F$) and incoherent features (electron correlation induced
localized band) are not observed in the up spin channel. The down
spin DOS, however, splits into two distinct features presumably
representing the incoherent features. Again, the signature of the
coherent feature at $\epsilon_F$ is absent. This is different from
that observed in the whole family of various 3$d$ and 4$d$
transition metal oxides.\cite{fujimori-review}

Interestingly, the gap between the incoherent features in the down
spin channel is significantly smaller than the gap in the up spin
channel. Thus, the electronic conduction is spin polarized even in
the insulating phase. Such {\it polar insulators} have immense
possibilities in technological applications involving spin-polarized
conduction such as magnetic sensors, quantum computations etc. The
gap between the incoherent features increases with the increase in
$U$ along with a redistribution of spectral weight among different
features.

In order to provide a realistic estimate of $U$ in this system, we
have calculated the experimental spectra as follows. The spin
integrated Ir 5$d$ density of states above -2 eV energies are
convoluted by the Fermi-Dirac distribution function and a Gaussian
representing the experimental resolution of 0.3 eV in the $x$-ray
photoemission spectroscopy.\cite{y2ir2o7} The calculated spectral
functions are shown by solid lines superimposed on the bulk spectra
obtained experimentally at 20~K; the comparisons are shown in Fig.
4(d). The result corresponding to the uncorrelated DOS is also shown
by dashed line in the figure.

The calculated spectrum corresponding to LSDA DOS is significantly
different from the experimental bulk spectrum. While the LSDA DOS
peaks at about 0.5 eV and extends down to about -1 eV, the intensity
in this energy range of the experimental spectrum is significantly
weak. The peak in the experimental spectrum appears at about 1.2 eV
and it extends down to about -2 eV. This suggests that the feature
around 1.2 eV represents the electronic states essentially localized
due to the electron correlation and can be termed as incoherent
feature. This also indicates that the electron correlation strength
may be about 2 - 3 eV. This is indeed large considering highly
extended nature of Ir 5$d$ orbitals and an estimation of smaller $U$
($\sim$ 0.6 eV) in 4$d$ transition metal oxides.\cite{srruo3}

\begin{figure}
\vspace{-4ex}
\begin{center}
\includegraphics[angle=0,width=0.5\textwidth]{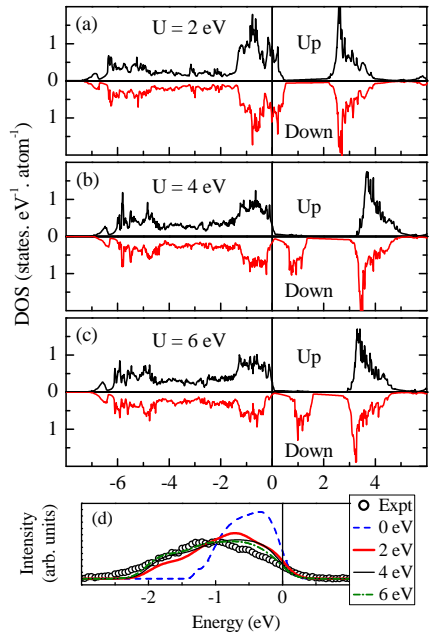}
\vspace{-2ex}
\end{center}
\caption{(Color online) Up and down spin density of states
calculated for different values of $U$ equal to (a) 2 eV, (b) 4 eV
and (c) 6 eV. (d) The calculated spin integrated spectral functions
for different values of $U$ is compared with the experimental bulk
spectrum at 20 K adapted from Ref.\cite{y2ir2o7}}
\end{figure}

Interestingly, the calculated results corresponding to finite $U$
values provide a good representation of the experimental results. In
addition to the resolution broadening, the experimental spectral
lineshape depends strongly on the transition matrix elements between
the ground state and final states, hole and electron lifetime
broadening etc, which are not considered in the present simulation.
Thus, the good correspondence of the calculated and experimental
results shown in Fig. 4(d) is remarkable. The change in lineshape is
marginal for $U \geq 2$~eV. These results clearly indicate that the
electron correlation is strong among Ir 5$d$ electrons. $U$ seem to
be comparable or higher than the estimations for 4$d$ transition
metal oxides.\cite{srruo3,ruthPRBR,ruti-PRB} Such anomaly may be
attributed to the narrow dispersion of the Ir 5$d$ bands leading to
significant local character in the corresponding electronic states.

The above results reveal another important point. The LSDA+$U$
results shown in Fig. 4 does not conform to the scenario of the
three peak structure (lower Hubbard band, coherent feature and the
upper Hubbard band) and its evolution with $U$ (spectral weight
transfers from the coherent feature to the incoherent feature with
the increase in $U$ but the intensity of the coherent feature at
$\epsilon_F$ remains unchanged) observed in correlated electron
systems. Evolution of up and down spin density of states are quite
different from each other and both are different from the above
Mott-Hubbard scenario widely discussed in the context of the
electronic properties of 3$d$ and 4$d$ transition metal oxides.
Since the electron correlation is not treated dynamically in
LSDA+$U$ calculations, which is important to capture such scenario,
the differences in the calculated results are not unusual. However,
the correspondence of the LSDA+$U$ results with the experimental
spectra is remarkable. Such consistency in a system, where various
approximations are expected to be good due to large radial extension
of the 5$d$ orbitals, suggests rethinking of the manifestation of
the correlation induced effects in the electronic structure.

Evidently no hard gap is found for $U$ as large as 6 eV. This is
consistent with the experimental observation of large spectral
intensity in the bulk spectra of this compound even at low
temperatures. This suggests that the insulating behavior in
Y$_2$Ir$_2$O$_7$ may not be due to a gapped phase. The intensity of
the spectral functions at $\epsilon_F$ is significantly weak. Thus,
localization effect due to defects, disorder and/or collective
excitation processes involving density waves {\it etc.} may be
important to determine the insulating phase in this
system.\cite{altshuler-aronov}

\section{Conclusions}

In summary, Y$_2$Ir$_2$O$_7$ is a unique system that exhibits weak
ferromagnetism and insulating transport. The small magnetic moment
in the paramagnetic phase indicates that intrinsic magnetic moment
is indeed weak, which is not unusual for highly extended 5$d$
orbitals. We have investigated the origin of such interesting
magnetic and electronic properties using band structure
calculations. The emphasis is on the role of spin-orbit coupling and
electron correlation in the electronic structure of this system.

The experimentally observed ferromagnetic phase could not be
captured using LSDA calculations presented here. The consideration
of spin-orbit coupling helps to reduce the magnetic moment, thereby,
brings the calculated magnetic moment closer to the experimentally
observed value. In addition, it lifts the degeneracy of the energy
bands. However, the experimental photoemission spectral functions
are very different from the LSDA results obtained with or without
spin-orbit coupling.

LSDA+$U$ results suggest that the electron correlation strength
required to derive the photoemission bulk spectra is significantly
large in contrast to the expectations in these systems. Narrow band
dispersion associated to the typical pyrochlore structure may be the
reason for such strong correlation. These results clearly suggest
rethinking of the role of spin-orbit coupling and electron
correlation in the electronic properties of oxides containing heavy
elements. It is evident that one needs to go beyond LSDA and
LSDA+$U$ approaches to determine the electronic structure of this
system. We believe that these results will provide significant
impetus to initiate studies of the correlated electron systems in
new directions.

\end{document}